# Nanomechanical detection of antibiotic-mucopeptide binding in a model for superbug drug resistance

JOSEPH WAFULA NDIEYIRA[1,2#], MOYU WATARI[1#], ALEJANDRA DONOSO BARRERA[1], DEJIAN ZHOU[3,4], MANUEL VÖGTLI[1], MATTHEW BATCHELOR[3], MATTHEW A. COOPER[5], TORSTEN STRUNZ[1], MIKE A. HORTON[1], CHRIS ABELL[3], TREVOR RAYMENT[6], GABRIEL AEPPLI[1] AND RACHEL A. McKENDRY[1*].

[1.] London Centre for Nanotechnology and Departments of Medicine and Physics, University College London, 17-19 Gordon Street, London WC1H 0AH, UK
[2.] Jomo Kenyatta University of Agriculture and Technology, Department of Chemistry, PO Box 62000, Nairobi, Kenya
[3.] Department of Chemistry, Lensfield Road, University of Cambridge, Cambridge CB2 1EW, UK
[4.] School of Chemistry and Astbury Centre for Structural Molecular Biology, University of Leeds, Leeds LS2 9JT, UK
[5.] Institute for Molecular Bioscience, The University of Queensland, Brisbane, 4072, Australia.
[6.] School of Chemistry, University of Birmingham, Edgbaston, Birmingham B15 2TT, UK
[#] These authors contributed equally
[*] email: R.A.McKendry@ucl.ac.uk

(Dated: 12 October 2008)

The alarming growth of the antibiotic-resistant superbugs methicillin-resistant Staphylococcus aureus (MRSA) and vancomycin-resistant Enterococcus (VRE) is driving the development of new technologies to investigate antibiotics and their modes of action. We report the label-free detection of vancomycin binding to bacterial cell wall precursor analogues (mucopeptides) on cantilever arrays, with 10 nM sensitivity and at clinically relevant concentrations in blood serum. Differential measurements quantified binding constants for vancomycin-sensitive and vancomycin-resistant mucopeptide analogues. Moreover, by systematically modifying the mucopeptide density we gain new insights into the origin of surface stress. We propose that stress is a product of a local chemical binding factor and a geometrical factor describing the mechanical connectivity of regions affected by local binding in terms of a percolation process. Our findings place BioMEMS devices in a new class of percolative systems. The percolation concept will underpin the design of devices and coatings to significantly lower the drug detection limit and may also impact on our understanding of antibiotic drug action in bacteria.

When biochemically specific interactions occur between a ligand immobilized on one side of a cantilever and a receptor in solution, the cantilever bends due to a change in surface stress[1-9]. The general applicability of this novel nanomechanical biosensing transduction mechanism has been shown for sequence-specific DNA hybridization[1-5,8], single base mismatches[1], DNA quadruplex[5], protein recognition[1,3,7,9] and recently the detection of interferon alpha induced I-8U gene expression in total human RNA, a potential marker for melanoma progression and viral infections[8]. However, to date, multiple cantilever arrays have not been applied to quantify drug-target binding interactions, despite offering considerable advantages. First, cantilevers require no reporter 'tags' or external probes and so biomolecules can be detected rapidly in a single step reaction. Second, cantilever arrays can screen multiple drug-target interactions and reference coatings in parallel and under identical experimental conditions. Third, we have previously shown that quantitative ligand-receptor binding constants can be measured on cantilever arrays[2]. Moreover, the nanomechanical signal is not limited by mass, in contrast to evanescent techniques such as surface plasmon resonance (SPR) that detects mass-related changes in the dielectric constant[10-13]. Cantilevers are therefore unique as probes of small molecule drug binding interactions and, by virtue of their miniaturized dimensions they are amenable for parallelization[14,15] for high-throughput screening of thousands of drugs per hour.

Here we report the first quantitative differential nanomechanical investigation of drug-target binding interactions on multiple cantilever arrays focusing on the antibiotic vancomycin (Figure 1). Today vancomycin is one of the last powerful antibiotics in the battle against resistant bacteria and the 'hospital superbug' MRSA[16-27]. It is a vital therapeutic drug used worldwide for the treatment of infections with Gram-positive bacteria, particularly those Staphylococci and Enterococci responsible for post-surgical infections. Vancomycin binds to the C-terminus of peptidoglycan mucopeptide precursors terminating in the sequence Lysine-D-Alanine-D-Alanine[16-18,20,21] as shown in Figure 1. This interaction blocks the action of bacterial transpeptidases and transglycosylases, which catalyse the cross-linking of the growing bacterial cell wall, resulting in cell lysis[16-27]. Unfortunately, due to the over-use of antibiotics, resistance to vancomycin is rapidly increasing and now poses a major international public health problem[22, 24,26]. Bacterial resistance in Enterococci can arise due to the subtle change of an amide linkage to an ester linkage in the growing bacterial cell wall, resulting in the deletion of a single hydrogen bond from the binding pocket, rendering the antibiotic therapeutically ineffective[16-27] (Figure 1). The development of novel methods to detect and quantify the binding of antibiotic–mucopeptide interactions is thus of significant clinical importance. In addition the structure and binding properties of vancomycin-mucopeptide complexes have been extensively studied both at surfaces and in free solution[16-27] and thus serve as a model system to evaluate the capabilities of cantilevers in small molecule drug-target detection.

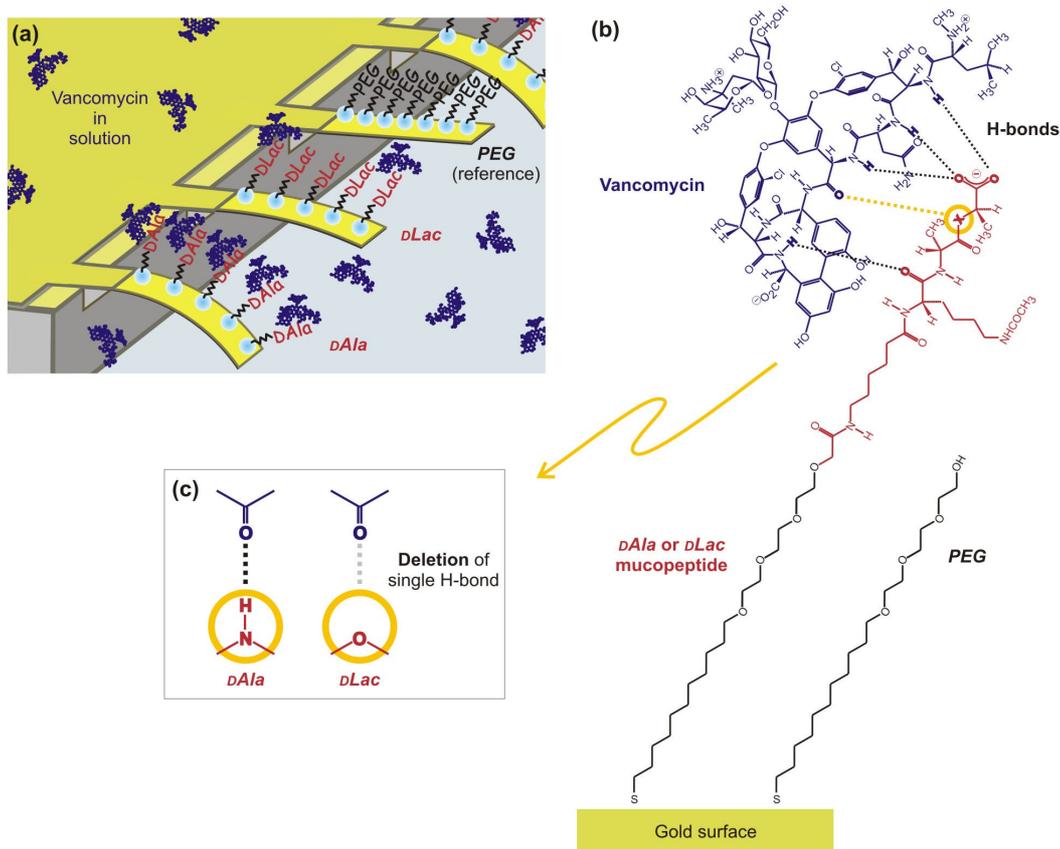

**Figure 1** The nanomechanical detection of vancomycin-mucopeptide analogue interactions on multiple cantilever arrays. (a) Schematic diagram to show cantilevers coated with ***DAla*** (vancomycin sensitive), ***DLac*** (vancomycin resistant) or ***PEG*** (reference) alkanethiol SAMs. Vancomycin is injected in solution and binds specifically to the mucopeptide analogues causing the cantilever to bend downwards due to a compressive surface stress. (b) The chemical binding interaction between vancomycin and the bacterial mucopeptide analogue, ***DAla***. It is known from solution phase studies that the specificity of this complex arises due to (i) the interaction of the C-terminal free carboxylate of the peptide with three amide bonds in the vancomycin backbone (ii) the formation of two C=O----H-N hydrogen bonds and (iii) hydrophobic interactions of the alanine methyl groups with aromatic residues of vancomycin. The dashed lines represent the 5 intermolecular hydrogen bonds. The yellow dashed line represents the hydrogen bond associated with bacterial resistance; (c) The deletion of a single H bond in mutated ***DLac*** mucopeptides gives rise to drug resistance. The binding pocket of vancomycin is represented schematically and the grey dotted line represents the deleted hydrogen bond and electrostatic repulsion between the oxygen lone pairs of electrons.

DETECTION OF VANCOMYCIN-MUCOPEPTIDE INTERACTIONS

To probe the in-plane nanomechanics of antibiotic drug-target interactions, multiple arrays of eight rectangular silicon cantilevers were coated on one side with a thin film of gold and functionalized with alkanethiol self-assembled monolayers (SAMs) of (i) the drug-sensitive mucopeptide analogue $HS(CH_2)_{11}(OCH_2CH_2)_3O(CH_2)(CO)NH(CH_2)_5(CO)$ -L-Lys-(ε-Ac)-D-Ala-D-Ala, herein termed ***DAla***; (ii) the mutated sequence which confers vancomycin resistance in VanA and VanB resistant Enterococcal phenotypes, $HS(CH_2)_{11}(OCH_2CH_2)_3O(CH_2)(CO)NH(CH_2)_5(CO)$-L-Lys-(ε-Ac)-D-Ala-D-Lac, termed ***DLac***. Our previous studies[2,5,6,8] have emphasized the importance of acquiring differential measurements using a reference cantilever and here we use a cantilever coated with an 'inert' SAM terminating in triethylene glycol $HS(CH_2)_{11}(OCH_2CH_2)_3OH$ termed ***PEG***, which is known to resist biomolecule adsorption on surfaces[28-30]. The Supplemental Material describes the synthesis of ***DAla***, ***DLac*** and ***PEG***. The absolute deflection at the free-end of each cantilever $\Delta z_{abs}$ was measured using a time-multiplexed optical detection system in different buffer and blood serum environments under constant flow. The bending signal was subsequently converted into a differential surface stress between the upper and lower sides of the

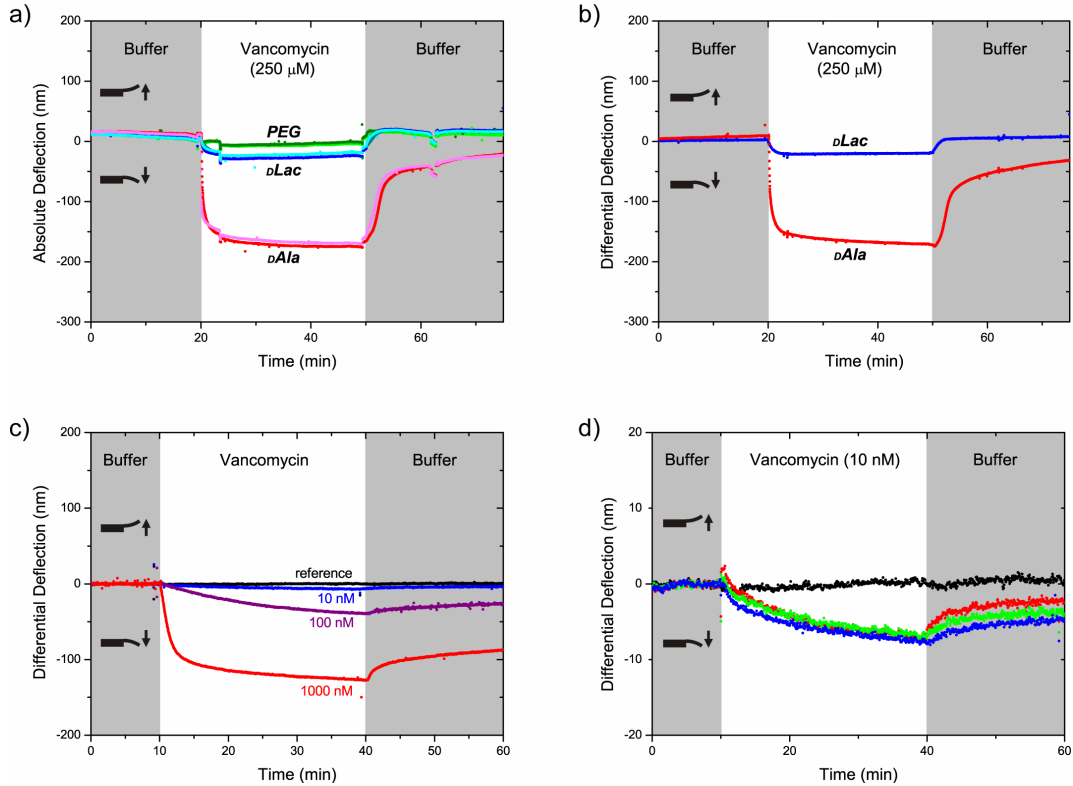

**Figure 2** Investigating the specificty and sensitivity of antibiotic-mucopeptide interactions on cantilever arrays. (a) Absolute bending signal of ***DAla₁*** (red), ***DAla₂*** (orange), ***DLac₁*** (light blue), ***DLac₂*** (dark blue), ***DLac₃*** (dark green) and in-situ reference cantilevers ***PEG₁*** (black) and ***PEG₂*** (grey) coated cantilevers to phosphate buffer, 250 μM vancomycin, and return to phosphate buffer. A negative signal corresponds to a compressive surface stress and the downwards bending of the cantilever, as illustrated in Figure 1a; (b) The corresponding differential bending signals of ***DAla₁*** (***DAla₁*** minus ***PEG₁***, shown in red) and ***DLac₁*** (***DLac₁*** minus ***PEG₁***, shown in blue); (c) Differential ***DAla*** signals for 10, 100, 1000 nM vancomycin. The differential ***PEG*** reference signal is shown (***PEG₂*** − ***PEG₁*** black); (d) Differential signals of three ***DAla*** cantilevers for 10 nM vancomycin. The differential ***PEG*** reference signal is shown (black).

cantilever $\Delta\sigma_{abs}$, using Stoney's equation[31]

$$\Delta\sigma_{abs} = \frac{1}{3}\left(\frac{t}{L}\right)^2 \frac{E}{1-\nu} \Delta z_{abs} \quad (1)$$

where $L$ is the effective length of the cantilever up to 500 μm, $t$ is the thickness ~ 0.9 μm, and $E/(1-\nu) = 180$ GPa is the ratio between the Young's modulus $E$ and Poisson ratio $\nu$ of Si (100)[32]. We used a home-built gravity flow system to control the exchange of up to six different vancomycin solutions (10 nM – 1 mM), 100 mM sodium phosphate buffer and 10 mM HCl regeneration solutions, via an automated valve (Supplemental Material).

The aim of our investigations was to ascertain whether cantilever arrays have the sensitivity to quantify vancomycin–***DAla*** binding interactions and detect the deletion of a single hydrogen bond associated with antibiotic resistance to the mutated peptide analogue, ***DLac***. Moreover, we examined the sensitivity of antibiotic detection in blood serum at clinically relevant concentrations[33] of 5-40 μg/ml, which corresponds to 3-27 μM. In addition, we sought to alter the surface peptide density in order to optimize drug detection sensitivity and to investigate the underlying mechanotransduction mechanism. Nanomechanical biosensors can best be exploited only if we develop a fundamental understanding of what causes the cantilever to bend and this knowledge will aid the development of new devices with significantly improved drug detection sensitivity.

The nanomechanical force exerted by vancomycin-peptide interactions was investigated on microfabricated cantilevers. The deflection of an array of cantilevers coated with ***DAla, DLac*** or in-situ reference ***PEG*** SAMs, was monitored in parallel upon injection of different concentrations of vancomycin in sodium phosphate buffer (pH 7.4, 0.1 M). Typical absolute bending signals for one array comprising two ***DAla***, three ***DLac*** and two ***PEG*** coated cantilevers are shown in Figure 2a. In buffer, we observed that all cantilevers showed a stable baseline. Upon injection of 250 μM vancomycin, both ***DAla₁*** and ***DAla₂*** coated cantilever rapidly bent downwards (illustrated in Figure 1a), reaching a stable equilibrium absolute compressive bending signal of -180 nm and -172 nm respectively after 30 mins under constant flow conditions. Conversely the ***DLac*** cantilevers showed a small absolute downwards bending signal of -38, -31 and -31 nm for ***DLac₁***, ***DLac₂*** and ***DLac₃***, respectively. The two reference ***PEG*** coated cantilevers, ***PEG₁*** and ***PEG₂*** showed a small downwards bending signal of -14 nm and -13 nm. Upon injection of the buffer, the signals were observed to converge back towards the stable 'zero stress' baseline. This step was then followed by 10 mM HCl, which is known to dissociate vancomycin-peptide complexes and regenerate the free peptides for further antibiotic studies.[13]

It is known that the absolute bending signals are a convolution of biologically *specific* binding events and *non-*

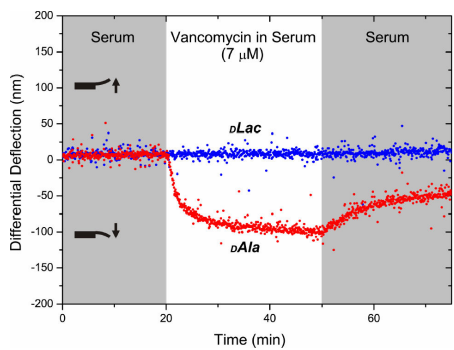

**Figure 3 Nanomechanical detection of antibiotics in blood serum at clinically relevant concentrations.** Differential bending signal of ***DAla₁*** (red) and ***DLac₁*** (blue) upon injection of 7 μM vancomycin in 90 % fetal calf serum and 10% sodium phosphate buffer at pH 7.4.

*specific* influences, including reactions occurring on the underside of the cantilever, liquid injection spikes, changes in refractive index and temperature[1,2,5-8]. These non-specific effects will affect ***DAla, DLac*** and ***PEG*** signals to the same extent, and are therefore removed by taking a differential measurement using a reference cantilever with an inert coating.[2,5-8] The differential measurements, shown in Figure 2b revealed the surface forces induced by biochemically *specific* vancomycin-peptide interactions. Upon injection of 250 μM vancomycin, the differential surface stress signal for ***DAla₁*** (*DAla₁ - PEG₁*) and ***DLac₁*** (*DLac₁ - PEG₁*) were found to be -35.3 mN/m and -5.1 mN/m respectively.

To examine the reproducibility of the nanomechanical signals, we performed more than 100 measurements on four different cantilever arrays, each composed of eight cantilevers. The raw bending signals were analyzed using automated data software developed in our group to rapidly examine large data sets and remove bias[6]. The average nanomechanical surface stress signal for 250 μM vancomycin on one array was -34.6 ± 0.9 mN/m for ***DAla*** and -4.2 ± 0.5 mN/m for ***DLac***. The mean signals across four arrays was -34.2 ± 5.9 mN/m for ***DAla*** and -3.8 ± 1.5 mN/m for ***DLac***. The high reproducibility of *within-array* measurements, and an increased variance associated with *between-array* measurements, agrees with our previous findings[6].

The dynamic range and sensitivity was investigated by varying the vancomycin concentration in solution [Van]. The differential ***DAla*** bending signal scaled with increasing [Van] ~ 10, 100 and 1000 nM, giving equilibrium differential signals of -8, -29 and -114 nm respectively (Figure 2c). The lowest [Van] to be detected was 10 nM giving rise to a ***DAla*** differential bending signal of -9 ± 2 nm on three cantilevers (Figure 2d).

The capacity of cantilever to detect antibiotics in serum was investigated in the clinically relevant concentration range[33] of 3-27 μM. Figure 3 shows the differential signal for ***DAla*** and ***DLac*** coated cantilevers upon injection of 7 μM vancomycin in serum (90 % fetal calf serum plus 10 % sodium phosphate buffer pH 7.4). The differential signal for ***DAla*** in serum was 105 ± 4 nm and no significant bending was detected for ***DLac***.

## DRUG-TARGET NANOMECHANICAL PERCOLATION

We monitored the nanomechanical response of cantilevers with systematically varied peptide densities *p* (where *p* is

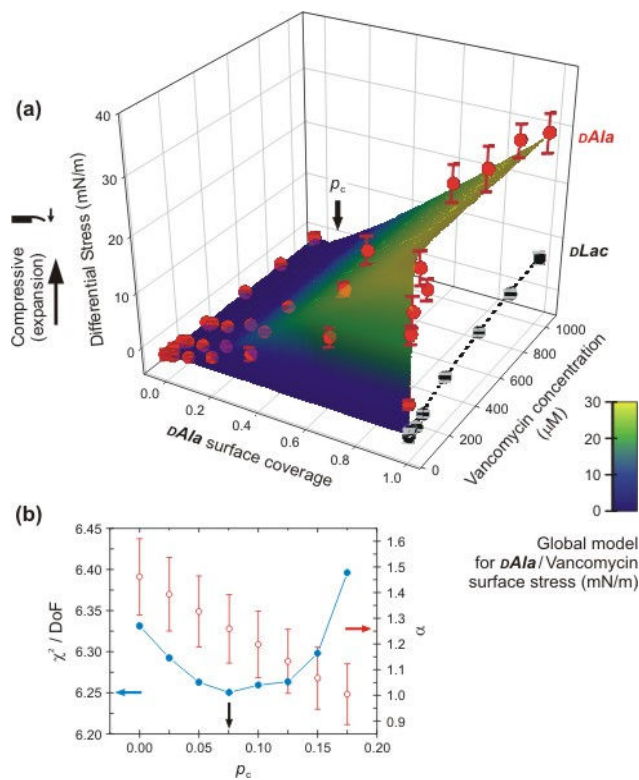

**Figure 4 Nanomechanical drug-target percolation on cantilever arrays.** (a) A three dimensional graph showing the measured differential surface stress response for ***DAla*** (red circles) and ***DLac*** (black circles) coated cantilevers as a function of vancomycin concentration in solution [Van] and ***DAla*** surface coverage p, superimposed with the results of the global fit according to Equation 2; (b) Least-squares analysis to determine the best values for $p_c$ and α. Plot shows the chi squared values (blue) and α (red) for each value of $p_c$. The black, blue, and red arrows indicate the chi squared minimum, $p_c$, and α values at the chi-squared minimum, respectively.

defined as the ***DAla*** surface coverage fraction on the cantilever determined by X-ray photoelectron spectroscopy (XPS) as described in the Supplemental Materials) to a series of [Van]. Figure 4 shows all of the stress data as a function of *p* and [Van] for both ***DAla*** and ***DLac***. The nanomechanical signal was much larger for ***DAla*** than ***DLac***, and steeply increased as a function of [Van] followed by saturation, whereas it increased more gradually as a function of *p*. For fixed *p* = 1, where the cantilever is coated with only ***DAla,*** the nanomechanical response saturates for [Van] > 50 μM when most accessible vancomycin binding sites are occupied, consistent with binding equilibrium. This effect is a measure of the specific chemical interactions between the vancomycin and the peptide. On the other hand, for fixed [Van], upon increasing ***DAla*** from *p* = 0 to *p* = 0.1 no detectable nanomechanical signal was measured, whereas from *p* = 0.1 to *p* = 1.0 there is an approximately linear increase. What this means is that the stress transduction is actually a *collective phenomenon*, requiring a relatively large fraction of the surface to be covered so as to establish connectivity between chemically transformed regions of the surface. Assuming that the local chemistry and geometric effects responsible for the

**Table 1** Equilibrium dissociation constant $K_d$ and saturation stress signals *a* of vancomycin-mucopeptide interactions on cantilever arrays compared with SPR and solution-phase UV spectroscopy measurements from the literature.

| Mucopeptide | *a* (mN/m) | $K_d$ cantilever (µM) | $K_d$ SPR (µM) | $K_d$ solution (µM) |
|---|---|---|---|---|
| *DAla* | 29.7 ± 1.0 | 1.0 ± 0.3 | 1.1 ± 0.1 (ref. 13) | 0.7 ± 0.1 (ref. 21) |
| *DLac* | 14.1 ± 3.0 | 800 ± 310 | 526 ± 139 (ref. 13) | 660 (ref. 25) |

The values stated are mean ± standard error for more than 100 cantilever measurements. Literature data also quoted in these units. No error analysis is given in ref. 11.

collective build-up of strain are separable, we can write a general product form for the cantilever response,

$$\Delta\sigma_{eq} = \frac{a \cdot [\text{Van}]}{K_d + [\text{Van}]} \left(\frac{p-p_c}{1-p_c}\right)^\alpha, \text{ for } p > p_c. \quad (2)$$

and zero otherwise. The first term is the Langmuir Adsorption Isotherm, accounting for drug-target binding events and the second term is the power law form describing the large-scale mechanical consequences of stressed network formation. The constant *a* corresponds to maximum surface stress when all the accessible binding sites are occupied and $K_d$ is the surface equilibrium dissociation constant on the cantilever. The build-up of surface stress follows from the connectivity of the chemically transformed network as well as the interactions between nodes of the network[34]. The exponent of the power law $\alpha$ is associated with the elastic interactions between chemically reacted regions on the cantilever. For short range interactions, such as steric neighbour-neighbour repulsive interactions there will be a finite percolation threshold $p_c$ beyond which there will be a connected network which can produce an apparent bending of the cantilever. On the other hand, for long range interactions such as ideally elastic interactions, $p_c=0$.

We have carried out a series of least-squares fits of Equation 2 to our data to find the key parameters as well as to ascertain the validity of the product form. The parameters $p_c$ and $\alpha$, which characterize the collective behavior are coupled in a statistical sense. Therefore, to determine what their best values might be, we have chosen not to rely on the multi-parameter fitting routine but instead have examined directly how the fit changes as a function of $p_c$, looking at both the resulting values of $\alpha$ and squared deviation between the data and the fit. Figure 4b shows the outcome, which reveals a percolation threshold $p_c = 0.075$, and a concomitant preference for a power $\alpha$ close to 1.3. We were able to ascertain the validity of the product form, Equation 2, by comparing the global fit values for $K_d$ and *a*, or $\alpha$ and $p_c$, with the values obtained from subsets of the data at constant *p* or constant [Van]. The analysis shown in the Supplemental Material shows that the values do not vary outside the experimental error limits, and that therefore, within the errors of our present experiments, local drug-target chemical interaction effects decouple from the collective elastic phenomenon ultimately responsible for the bending of the entire cantilever. Having established this, we can find $K_d$, *a*, $\alpha$, and $p_c$ using all available data. The outcome of the global fit is superimposed onto the measured differential stress signals in Figure 4a. Table 1 shows a summary of *a* and $K_d$ for both *DAla* and *DLac* peptides and makes the comparison with reported values from solution phase UV spectroscopy[20,21] and SPR measurements[11-13].

## CONCLUSIONS

Our experiments show that cantilever arrays have the sensitivity to detect and quantify the binding affinity of the antibiotic vancomycin to drug-target mucopeptide analogues: Lysine-D-Alanine-D-Alanine and Lysine-D-Alanine-D-Lac. The former occurs in the peptidoglycan precursors found in vancomycin sensitive bacteria and the latter in those precursors in VanA and VanB vancomycin resistant bacteria[16-27]. Differential measurements could successfully discriminate between the two peptide sequences, detecting the deletion of a single hydrogen bond from the drug binding pocket, which is associated with drug resistance. This gave rise to an 800-fold increase in $K_d$ of *DLac* compared to *DAla*, in agreement with measurements made by SPR[11-13].

We find that the minimum detectable vancomycin concentration was 10 nM and comparable to SPR studies, which have reported the detection of 30 nM vancomycin[11-13]. Furthermore, we show that cantilevers can detect and quantify vancomycin in blood serum at clinically relevant concentrations, which is important for pharmacokinetic/dynamic drug profiling, personalized medicine and forensic applications.

The molecular binding events occurring between vancomycin in solution and *DAla* were found to generate a repulsive compressive surface stress. The origin of the biochemically induced surface stress is the subject of much scientific debate and interest[1-9]. Our experiments reveal a finite percolation threshold $p_c = 0.075$ below which the macroscopic bending is effectively zero. This means that a critical number of *DAla* and vancomycin binding events are required to yield observable stress and demonstrates a local short range transduction mechanism. For $p > p_c$, the nanomechanical signal increases where the exponent in the power law $\alpha$ is 1.3 and so is approximately linearly proportional to the number of *DAla* molecules on the cantilever. Figure 5 shows the operation of this mechanism, which begins with the steric forces generated by insertion of the vancomycin into the *DAla* SAM. The resulting complexes will induce a local strain in the silicon as well as carry an electrostatic charge, which in the neutral pH conditions of this study is +1 for vancomycin. As the number of such regions grows they will interact to produce bending of the entire cantilever.

The proposed percolative mechanisms of antibiotic-mucopeptide triggered changes in surface stress differ significantly from previous studies of the Young's modulus of macroscopic two-dimensional model random elastic media[35,36]. Here we establish nanomechanical cantilever biosensors as the hosts for new universality classes of percolating systems, where the elasticity resides in the coupled multilayered, multiscale system[37] - buffer/Van/*DAla*/*PEG*/gold/chromium/silicon. Our findings suggest that the structure and mechanics of the underlying *DAla*/*PEG* SAMs play a major role in mechanotransduction. Future work will investigate the extent of mixed monolayer phase separation[38,39] on cantilevers and the decoupling of chemical and geometric factors (manuscript in preparation MW, RAM & GA). These findings will aid rational design of novel devices and surface chemistries for improving the sensitivity of cantilevers to chemical binding events such as those in our current drug testing application. Interestingly we show that the maximum stress signal is obtained at high *DAla* packing densities, conditions that are traditionally considered to be unfavorable for other surface sensing techniques such as SPR[12, 13] (For p=1.0 we estimate $10^{11}$ ***DAla*** per cantilever, with a single footprint of 44 Å$^2$, as described in the XPS measurements detailed in the Supplemental Material). Thus, beyond our key result that cantilevers are a very useful tool in antibiotic research, our systematic experiments on model SAMs provide a new framework for understanding and eventually engineering the response of cantilevers to biochemical signals.

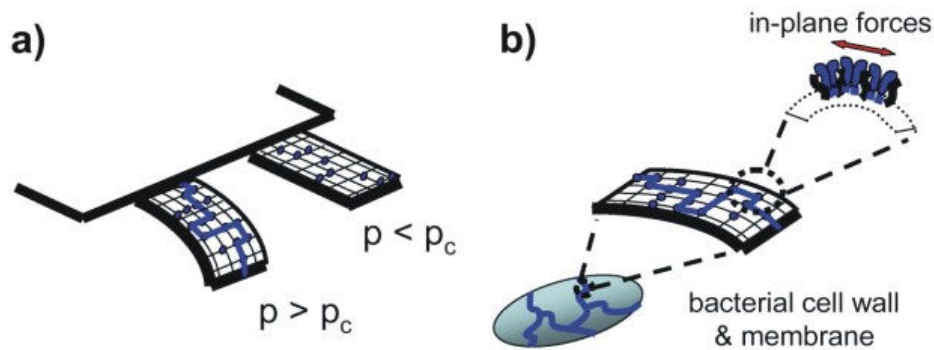

**Figure 5** Concepts underpinning nanomechanical antibiotic transduction. **(a)** A schematic to show the concept of percolation on a cantilever array; **(b)** A schematic to illustrate nanomechanical drug-target percolation on a bacterial membrane and cell wall.

We close by speculating that nanomechanical percolation may play an important role not only in sensor response but also in the glycopeptide antibiotic mode of action in real bacteria. Williams and co-workers proposed that glycopeptide antibiotic dimerisation impacts on antimicrobial efficacy and that antibiotic dimers show unusually strong binding constants[16-18]. We propose further that drug-target binding events act *collectively* to disrupt the mechanical properties of the bacterial peptidoglycan cell wall and plasma membrane (Figure 5b). Dimeric glycopeptides with membrane-insertive hydrophobic tails such as biphenylchloroeremomycin and oritavancin[18], may possess greater potency than vancomycin because of their propensity to self-associate and diffuse in the membrane to create extended networks. Our results should therefore clearly motivate future work to investigate the percolation of drug action in lipid membranes, including eventually those of living bacteria, and test the more general validity of the hypothesis (underlying Equation 2) of decoupling between chemical binding and geometric factors.

**Supplemental Material accompanying this paper is provided on the following pages.**

### Acknowledgements
We thank the Engineering and Physical Sciences Research Council (Speculative Engineering funding programme JWN, RMcK, GA, MW, DZ,), Interdisciplinary Research Centre in Nanotechnology (Cambridge, UCL and Bristol: MW, JWN, RMcK), the Royal Society (RMcK, JWN), Biotechnology and Biological Science Research Council (RMcK), the Wolfson Foundation (GA), Cancer Research UK (ADB) for funding. We thank Emily Smith (University of Nottingham EPSRC XPS Open Access scheme), Jane and Rex Galbraith (UCL Statistics) for helpful discussions.

### Author contributions
JWN and MW contributed equally to this manuscript. All authors discussed the results and commented on the manuscript.

### Author Information
Correspondance and requests for materials should be adressed to R.A.M.

# Supplemental Material

'Nanomechanical detection of antibiotic-mucopeptide binding in a model for superbug drug resistance'

J. W. Ndieyira, M. Watari, A. Donoso Barrera, D. Zhou, M. Vögtli, M. Batchelor, M. A. Cooper, T. Strunz, M. A. Horton, C. Abell, T. Rayment, G. Aeppli and R. A. Mckendry*.

*email: R.A.McKendry@ucl.ac.uk

**Synthesis of thiolated peptides**

The peptides *DAla* and *DLac* were synthesized by solid phase methodology using commercially available pre-loaded Wang-D-Ala and Wang-D-Lac resins and standard Fmoc-protecting group chemistry.[S1] The cleaved products were further purified by reverse phase HPLC by varying the mobile phase from 5% to 95% of acetonitrile in water (with 0.5% trifluoroacetic acid). The peptides were characterized by NMR and HRMS (+ESI, Q-TOF) as follows;

**i)** D-Ala-OH (HS(CH2)$_{11}$(EG)$_3$-OCH$_2$-Ahx-L-Lys-D-Ala-D-Ala-OH
$^1$H NMR (δ ppm, 400 mHz, CD$_3$OD): 1.25-1.80 (m, 36H, [15CH$_2$ + 2 CH$_3$]), 1.91 (s, 3H, CH$_3$), 2.24 (t, 2H, CH$_2$), 2.47 (t, 2H, J = 7.1 Hz, C$H_2$SH), 3.14 (t, 2H, J = 7.0 Hz, NHC$H_2$), 3.22 (t, J = 7.0 Hz, 2H, NHC$H_2$), 3.44 (t, 2H, J = 6.7 Hz, -C$H_2$(OEG)-), 3.54-3.69 (m, 12H, 3(EG)), 3.96 (s, 2H, OC$H_2$C=O), 4.21 (t, 1H, J = 7.1 Hz, L-lys-α-C$H$), 4.31-4.42 (m, 2H, 2 [D-Ala-α-C$H$]).
$^{13}$C NMR (δ ppm, 125 mHz, CDCl$_3$): 176.24, 175.58, 174.46, 174.29, 173.21, 172.61 (6 *C*=O), 72.39, 71.95, 71.55, 71.52, 71.37, 71.22, 71.16 (PEG), 55.15, 50.00, 40.17, 39.83, 36.50, 35.22, 32.27, 30.72, 30.70, 30.64, 30.56, 30.27, 30.21, 29.40, 27.57, 27.21, 26.46, 24.97, 24.23, 22.57, 18.01, 17.50.
HRMS (+ESI, Q-TOF), found 842.4941; required for C$_{39}$H$_{73}$N$_5$O$_{11}$SNa [M+Na]$^+$, 842.4925, dev. 1.86 ppm.

**ii)** D-Lac-OH (HS(CH2)$_{11}$(EG)$_3$-OCH$_2$-Ahx-L-Lys-D-Ala-D-Lac-OH
$^1$H NMR (δ ppm, 500 MHz, CD$_3$OD): 1.25-1.80 (m, 36H, [15CH$_2$ + 2CH$_3$]), 1.92 (s, 3H, CH$_3$), 2.24 (t, 2H, J = 7.0 Hz, CH$_2$), 2.47 (t, 2H, J = 7.0 Hz, C$H_2$SH), 3.13-3.17 (m, 2H, NHC$H_2$), 3.18-3.24 (m, 2H, NHC$H_2$), 3.46 (t, 2H, J = 6.7 Hz, -C$H_2$(OEG)), 3.55-3.68 (m, 12H, 3(EG)), 3.96 (s, 2H, OC$H_2$C=O), 4.30-4.40 (m, 1H, L-lys-α-C$H$), 4.40-4.50 (m, 1H, D-Ala -α-C$H$]). 5.00-5.10 (m, 1H, D-Lac-α-C$H$).
$^{13}$C NMR (δ ppm, 125 MHz, CD$_3$OD): 175.99, 174.12, 173.34, 173.20, 172.84, 172.60 (6 *C*=O), 79.47, 72.40, 71.99, 71.58, 71.55, 71.39, 71.24, 71.19 (PEG), 54.37, 53.68, 40.19, 39.82, 36.73, 36.63, 35.23, 32.84, 32.22, 30.74, 30.71, 30.64, 30.57, 30.26, 30.21, 30.00, 29.85, 29.41, 27.58, 27.52, 27.22, 26.62, 26.56, 24.98, 24.17, 22.60 (*C*H$_3$), 17.51 (*C*H$_3$), 17.44 (*C*H$_3$).
HRMS (+ESI, Q-TOF), found 843.4761; required for C$_{39}$H$_{73}$N$_5$O$_{11}$SNa [M+Na]$^+$, 843.4765, dev. – 0.53 ppm.

**iii)** The synthesis of the 11-mercaptoundecyl tri(ethylene glycol) alcohol (*PEG*) has been described.[S2] The crude compound was purified using flash column chromatography on silica (10% ethanol in ethyl acetate) to yield the final product (*PEG*) as a colourless oil. $^1$H NMR (250 MHz, CDCl$_3$, δppm): 1.20-1.37 (m, 14H, 7C$H_2$), 1.49-1.60 (m, 4H, 2C$H_2$), 2.46 (q, 2H, J = 7.0 Hz, HSC$H_2$-), 3.05 (s, br, 1H, -O$H$), 3.40 (t, 2H, J = 7.0 Hz, -C$H_2$PEG), 3.50-3.75 (m, 12H, 3(OC$H_2$C$H_2$). $^{13}$C NMR (62.5 MHz, CDCl$_3$, δppm): 72.5, 71.4, 70.5, 70.3, 69.9, 61.5, 34.0, 33.7,

30.5, 29.5, 29.4, 29.0, 28.8, 28.7, 28.3, 26.0, 24.5. HRMS (Q-TOF, ES$^+$), found 359.2222; required for C$_{17}$H$_{36}$O$_4$SNa [M + Na]$^+$, 359.2232.

**Preparation of cantilevers**

Cantilever arrays were fabricated by IBM Research Laboratory, Rüschlikon, Switzerland, and purchased from Veeco Instruments Inc. (Santa Barbara, CA, U.S.A.). Each Si (100) cantilever was nominally 500 μm long, 100 μm wide, and 0.9 μm thick. Cantilever arrays were first cleaned with freshly prepared piranha solution (at ratio 1:1 H$_2$SO$_4$ and H$_2$O$_2$) for 20 mins. Piranha solution had to be handled with care as it is hazardous and can cause explosions or severe skin burns. Arrays were then thoroughly rinsed in deionized water before being immersed in the second freshly prepared piranha solution for another 20 mins, and again rinsed thoroughly with deionized water. Finally, the arrays were rinsed with pure ethanol and dried on a hotplate at 75 °C. They were then inspected using the optical microscope to confirm their cleanliness before transferring to the evaporation chamber (BOC Edwards Auto 500, U.K.) for overnight pumping. One side of each cantilever array was coated by thermal evaporation with a 2 nm Cr adhesion layer followed by 20 nm of gold from a base pressure ~ 2×10$^{-7}$ mbar, and using evaporation rates of 0.02 nm/s for Cr and Au, as measured directly above the source by a quartz crystal monitor. Once the required thickness was attained, the samples were left in the chamber for 1-2 h to cool under vacuum before opening. The freshly evaporated cantilevers were sealed in a vacuum storage vessel (Agar Scientific, U.K.) filled with argon and functionalized within a few hours after the evaporation. The cantilevers were incubated in an array of eight glass microcapillaries filled in a random order with 2 mM ethanolic solutions of *DAla*, *DLac* or *PEG* for 20 mins, and then rinsed in pure ethanol and deionized water. The cantilevers were stored under distilled water immediately after functionalisation until use. Cantilevers were used immediately but where it was not possible to use all of them within a day, then they were kept in the Petri dishes with deionised water as we have previously found that functionalised arrays could be stored under deionised water for 2 weeks without degrading as opposed to being stored in air.

**Preparation of solutions**

Buffer solutions were prepared by mixing 0.1 M mono- and di-basic sodium phosphate salts (Sigma-Aldrich, U.K.) dissolved in ultrapure water (18.2 MΩ·cm resistivity, Millipore Co., Billerica, MA, U.S.A.) to yield a pH value of 7.4. Fresh buffer was used to prepare the solutions of vancomycin (Sigma-Aldrich, U.K). The solutions were filtered using 0.2 μm filters (Millipore) and ultrasonicated for 30 mins at room temperature before being purged with argon. To investigate the detection sensitivity of cantilevers to antibiotics in serum, vancomycin solutions were prepared using 90% fetal calf serum (Gibco, USA) with 10 % sodium phosphate buffer.

**Instrumentation**

The absolute bending of all eight cantilevers was monitored using the serial time multiplexed optical beam method with a single position sensitive detector (Scentris, Veeco Instruments Inc.). The functionalized cantilever array was mounted in a sealed liquid chamber with a volume of approximately 80 μl. The liquid cell and each aliquot of aqueous solutions were placed into the temperature controlled cabinet overnight to allow for temperature equilibration before undertaking each experiment. Alignment of each laser spot onto the free-end of each cantilever on the array was confirmed using 1 °C temperature heating test. The purpose of the heating test was to ensure that the effective length of all eight cantilevers was similar. This was witnessed by downward bending, due to the bimetallic effect. The relative standard deviation of the absolute bending signals of all eight cantilevers was ≤ 5%. Herein a negative deflection signal denotes a compressive surface stress and a positive signal is tensile. The efficient exchange of liquids in the liquid cell was achieved with a home-built gravity flow microfluidics system. The desired flow rate was first determined before each experiment to ensure a constant flow rate for all the arrays. The data acquisition was automated using LabView (National Instruments Co., Austin, TX, U.S.A.) software via a 6-way valve (Serial MVP, Hamilton, Reno, NV, U.S.A.). The measurement protocol involved the following: i) sodium phosphate buffer solution (pH 7.4, 0.1 M) for 30 mins to establish a baseline; ii) injection of vancomycin in sodium phosphate buffer for 30 mins; iii) to dissociate the complex we then injected sodium phosphate buffer (pH 7.4 , 0.1 M) wash for 30 mins; iv) a further washing step using 10 mM HCl for another 30 mins to remove the vancomycin and regenerate the peptide surface; v) finally another sodium phosphate buffer step to restore the baseline signal. This process was repeated for all the vancomycin concentrations. All signals were acquired under a constant liquid flow rate of 180 ± 30 μl/min.

**Validity of the product form of Equation 2**

Least square fits were applied to subsets of the data, in the form of constant p or constant [Van] cuts through the data, and determining whether $K_d$ and $a$ or $p_c$ and $\alpha$ vary outside the experimental error limits with [Van] and $p$, respectively. The results are summarized in Table S1. The fitted parameters $a$ = (29.7 ± 1.0) mN/m, $K_d$ = (1.0 ± 0.3) μM, $p_c$ = (0.08 ± 0.09), and $\alpha$ = (1.3 ± 0.3) given in the first row were obtained by including all available data; the fitted parameters $a$ = (29.5 ± 2.0) mN/m and $K_d$ = (0.8 ± 0.5) μM at fixed $p_c$ = 0.08 and $\alpha$ = 1.3 given in the second row were obtained by including $p$ = 1.0 data only; the fitted parameters $p_c$ = (0.00 ± 0.40) and $\alpha$ = (1.9 ± 1.3) at fixed $a$ = 29.7 mN/m and $K_d$ = 1.0 μM given in the third row were obtained by including [Van] = 10 μM data only; the fitted parameters $p_c$ = (0.09 ± 0.17) and $\alpha$ = (1.2 ± 0.5) at fixed $a$ = 29.7 mN/m and $K_d$ = 1.0 μM given in the fourth row were obtained by including [Van] = 100 μM data only; the fitted parameters $p_c$ = (0.17 ± 0.07) and $\alpha$ = (0.8 ± 0.3) at fixed $a$ = 29.7 mN/m and $K_d$ = 1.0 μM given in the fifth row were obtained by including [Van] = 250 μM data only. These results show that the fitted parameters $a$, $K_d$, and $\alpha$ obtained from the global fit and the fitted parameters obtained from the fit of data subsets were consistent, and that therefore, to within experimental error, local chemical interaction effects decouple from the collective elastic phenomenon ultimately responsible for the bending of the entire cantilever.

**Table S1**: Least square fits of data subsets to validate the product form of Equation 2.

| $p$ | [Van] (μM) | $a$ (mN/m) | $K_d$ (μM) | $p_c$ | $\alpha$ |
|---|---|---|---|---|---|
| all[a] | all[b] | 29.7 ± 1.0 | 1.0 ± 0.3 | 0.08 ± 0.09 | 1.3 ± 0.3 |
| 1.0 | All | 29.5 ± 2.0 | 0.8 ± 0.5 | 0.08 | 1.3 |
| all | 10 | 29.7 | 1.0 | 0.00 ± 0.40 | 1.9 ± 1.3 |
| all | 100 | 29.7 | 1.0 | 0.09 ± 0.17 | 1.2 ± 0.5 |
| all | 250 | 29.7 | 1.0 | 0.17 ± 0.07 | 0.8 ± 0.3 |

[a] *Including all values of p investigated herein, i.e. 0.05, 0.1, 0.3, 0.5, 0.7, 0.9, and 1.0.*
[b] *Including all values of [Van] investigated herein, i.e. 0.1, 1, 10, 50, 100, 250, 500, 750, 1000 μM*

## X-ray photoelectron spectroscopy

The relation between the solution molar fraction and surface coverage of *DAla* was measured on the resultant SAMs by X-ray photoelectron spectroscopy. Given the miniaturised geometry of cantilevers, characterisation of the SAMs was performed on silicon wafers instead, which had been functionalized in parallel with the silicon cantilevers. The data were captured using Kratos VISION II software (version 2.2.6) on a Kratos Axis Ultra spectrometer equipped with aluminum Al $K_\alpha$ source and 1486.6 eV line energy. The X-ray spot size was 1 mm² and the analysis area was defined by the SLOT aperture of 300 × 700 μm² with a hybrid (magnetic/electrostatic) optics and a multi-channel plate and delay line detector (*DLD*) with a collection angle of 30° and a take off angle of 90°. The pressure in the chamber was $3 \times 10^{-9}$ Torr. Preliminary wide scans were taken over the full range of 1400 to 0 eV, with step size of 1 eV and pass energy of 80 eV to improve signal to noise ratio on the peaks of interest and then reduced down to 700 to 0 eV for the main analysis of the high resolution scans on N (1s), S (2p), Au (4f), and C (1s) spectra acquired with a 20 eV pass energy using scan step size of 0.1 eV. The averaging was undertaken over three spectra in three separate areas on each sample.

Figure S1 shows typical raw high-resolution photoemission spectra of N (1s), S (2p), and Au (4f) acquired on a SAM prepared from the ethanolic solution at a *DAla* to *PEG* ratio of 0.50:0.50 superimposed to fitted curves. The peak fitting of the raw spectra was carried out using commercial software (CasaXPS, 2.3.2) by using the same line shape and full width half maximum on each sample, with a linear background subtraction. Errors in the fitted area under the photoemission peaks were calculated as the standard error of the mean of the difference between the raw and fitted data.

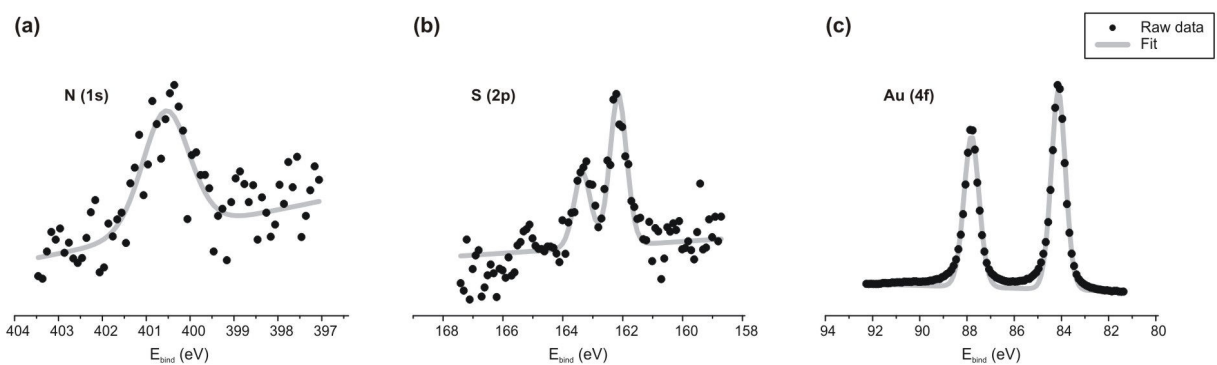

**Figure S1:** Typical raw high-resolution photoemission spectra of (a) N (1s), (b) S (2p), and (c) Au (4f) acquired on a SAM prepared from the ethanolic solution at a *DAla* to *PEG* ratio of 0.50:0.50, superimposed to fitted curves in grey.

## Langmuir adsorption models

The competitive adsorption of *DAla* and *PEG* thiols onto gold can be described by the chemical reaction

$$[\text{DAla}]^{\text{solution}} + [\text{PEG}]^{\text{surface}} \leftrightarrow [\text{DAla}]^{\text{surface}} + [\text{PEG}]^{\text{solution}}, \qquad \text{Equation S1}$$

where $[\text{DAla}]^{\text{solution}}$ and $[\text{PEG}]^{\text{solution}}$ is the solution concentration and $[\text{DAla}]^{\text{surface}}$ and $[\text{PEG}]^{\text{surface}}$ is the surface concentration of *DAla* and *PEG*, respectively. The equilibrium of this reaction is characterised by the equilibrium constant

$$K_{eq} = \frac{[\text{DAla}]^{\text{surface}} [\text{PEG}]^{\text{solution}}}{[\text{DAla}]^{\text{solution}} [\text{PEG}]^{\text{surface}}} . \qquad \text{Equation S2}$$

In the cantilever experiments presented herein, we varied the ratio of *DAla* and *PEG* thiol solutions at a fixed total solution concentration of thiols $[\text{thiol}]^{\text{solution}} = [\text{DAla}]^{\text{solution}} + [\text{PEG}]^{\text{solution}}$ of 2 mM (dissolved in ethanol). Therefore the solution molar fraction of *DAla* and *PEG* totals to $\chi_{\text{DAla}}^{\text{solution}} + \chi_{\text{PEG}}^{\text{solution}} = 1.0$. Similarly, the total surface concentration of the thiols is $[\text{thiol}]^{\text{surface}} = [\text{DAla}]^{\text{surface}} + [\text{PEG}]^{\text{surface}}$ and the surface coverage of the thiols is $\chi_{\text{thiol}}^{\text{surface}} = \chi_{\text{DAla}}^{\text{surface}} + \chi_{\text{PEG}}^{\text{surface}}$. The total surface coverage of the thiols is defined as $\chi_{\text{thiol}}^{\text{surface}} = 1.0$ if the surface coverage of *DAla* and *PEG* are described relative to one another, whereas $\chi_{\text{thiol}}^{\text{surface}}$ depends on $\chi_{\text{DAla}}^{\text{surface}}$ and $\chi_{\text{PEG}}^{\text{surface}}$ if the surface coverage of *DAla* and *PEG* are related to the total adsorption sites on the gold surface, which is due to differences in the molecular volume (hence steric effects) between *DAla* and *PEG* thiol molecules.

To describe the surface coverage of *DAla* as a function of its solution molar fraction, the binding constant of *DAla* $K_{\text{DAla}}$ is formulated by means of Equation S2 and the relations $\chi_{\text{DAla}}^{\text{solution}} + \chi_{\text{PEG}}^{\text{solution}} = 1.0$ and $\chi_{\text{thiol}}^{\text{surface}} = 1.0$, which is then solved for $\chi_{\text{DAla}}^{\text{surface}}$ to give

$$\chi_{\text{DAla}}^{\text{surface}}\left(\chi_{\text{DAla}}^{\text{solution}}, K_{\text{DAla}}\right) = \frac{K_{\text{DAla}} \chi_{\text{DAla}}^{\text{solution}}}{1 + \chi_{\text{DAla}}^{\text{solution}}(K_{\text{DAla}} - 1)} .$$

Equation S3

The total surface coverage of thiols $\chi_{\text{thiol}}^{\text{surface}}\left(\chi_{\text{DAla}}^{\text{surface}}, \chi_{\text{PEG}}^{\text{surface}}\right) = \chi_{\text{DAla}}^{\text{surface}} + \chi_{\text{PEG}}^{\text{surface}}$ can then be derived by combining Equation S3 with Equation S2 and solving the expression for the binding constant of thiols $K_{\text{thiol}}$ for $\chi_{\text{PEG}}^{\text{surface}}$, which reads

$$\chi_{\text{PEG}}^{\text{surface}}\left(\chi_{\text{DAla}}^{\text{solution}}, K_{\text{DAla}}, K_{\text{thiol}}\right) = \frac{K_{\text{DAla}}\left(1 - \chi_{\text{DAla}}^{\text{solution}}\right)}{K_{\text{thiol}}\left(1 + \chi_{\text{DAla}}^{\text{solution}}(K_{\text{DAla}} - 1)\right)}$$

Equation S4

and so the total surface coverage of thiols corresponds to the sum of Equations S3 and S4.

**Analysis of *DAla* and thiol surface coverage**

The surface coverage of *DAla* peptides $\chi_{\text{DAla}}^{\text{surface}}$ can be directly correlated with the intensities of N (1s) photoelectrons, since nitrogen is present in the molecular sequence specific to the *DAla* peptide only. The surface coverage of thiols $\chi_{\text{thiol}}^{\text{surface}}$ was measured as the ratio of the intensities of S (2p) and Au (4f) photoelectrons, which have the same escape depth and are thus attenuated to the same extent by the organic overlayer. Figure S2 shows the measured N (1s) intensities and the ratios of the intensities of S (2p) and Au (4f), both normalised to the 100% *DAla* SAM, as a function of *DAla* solution molar fraction and superimposed to the fit of Langmuir model Equation S2 and the sum of Equation S2 and S3, respectively. As can be seen in the figure, the relation between the measured surface coverage and the *DAla* solution molar fraction was well described by the Langmuir models. The fitted binding constants, $K_{\text{DAla}} = 0.25$ for the adsorption of *DAla* peptides and $K_{\text{thiol}} = 0.15$ for the adsorption of thiols suggest that the *DAla* peptides adsorbed slower and had a lower packing density compared to *PEG* molecules. For example, the thiol coverage of the 100% *DAla* SAM exhibited a 39% lower value compared to the 100% *PEG* SAM. This result is rather intuitive because of the larger molecular volume occupied by the *DAla* peptides as well as the increased molecular length, both giving rise to greater steric effects and diffusion length compared to more compact *PEG* molecules.

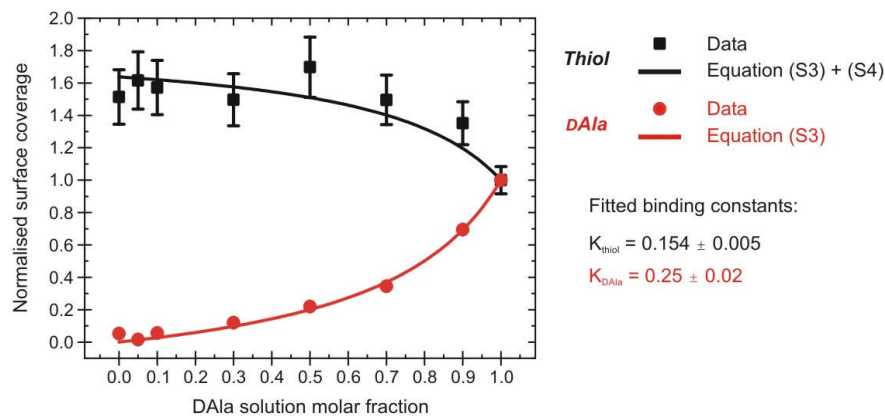

**Figure S2:** Normalised N (1s) intensities and the ratios of the intensities of S (2p) and Au (4f), as a function of *DAla* solution molar fraction and superimposed to the fit of Langmuir model.

The analysis of the surface coverage allows us to evaluate the cantilever measurements performed at different ***DAla*** peptide densities and vancomycin solution concentrations. Figure S3 shows the differential stress as a function of (a) the ***DAla*** surface coverage superimposed to the percolation model (i.e. geometrical term in Equation 2) and (b) the solution vancomycin concentration superimposed to the Langmuir model (i.e. chemical term in Equation 2). The three-dimensional representation of these data is presented in Figure 4 of the manuscript.

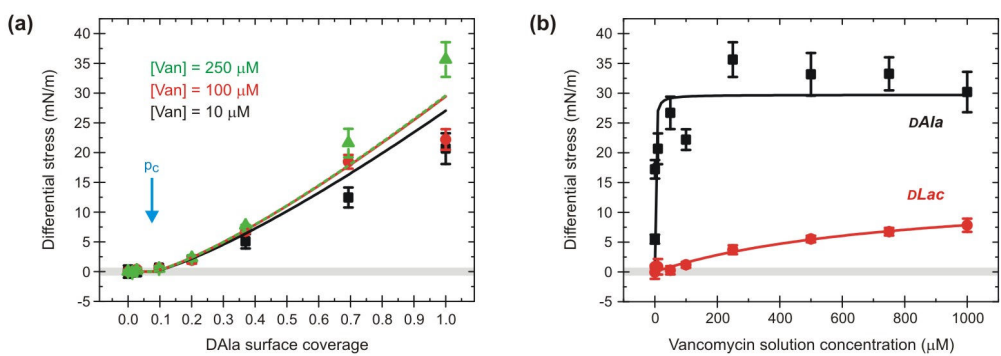

**Figure S3:** Measured differential surface stress response for ***DAla*** and ***DLac*** coated cantilevers as a function of (a) ***DAla*** surface coverage and (b) vancomycin concentration in solution [Van] superimposed with the results of the global fit according to Equation 2 (solid lines).

Furthermore, our X-ray photoelectron spectroscopy data can give an estimate of the surface area occupied by one ***DAla*** peptide molecule, by referring to the literature value of 27 Å$^2$ per thiolate reported for ***PEG*** SAMs.[S3] This translates into a molecular area of 44.2 Å$^2$ in the 100% ***DAla*** SAM or a nearest-neighbour distance of 7.1 Å by assuming an ideal hexagonal packing.